\documentclass[twocolumn,preprintnumbers,amsmath,amssymb]{revtex4}
\usepackage{graphicx}
\usepackage{amssymb}

\begin{document}

\title{Cryptanalysis and improvement of Wu-Cai-Wu-Zhang's quantum private
comparison protocol}

\author{Guang Ping He}
\email{hegp@mail.sysu.edu.cn}
\affiliation{School of Physics, Sun Yat-sen University, Guangzhou 510275, China}

\begin{abstract}
In a recent paper (\textit{Int. J. Quantum Inf.} \textbf{17} (2019)
1950026), the authors discussed the shortcomings in the security of a
quantum private comparison protocol that we previously proposed (\textit{Int. J. Quantum Inf.} \textbf{15} (2017) 1750014). They also
proposed a new protocol aimed to avoid these problems. Here we analysis the
information leaked in their protocol, and find that it is even less secure
than our protocol in certain cases. We further propose an improved version
which has the following advantages: (1) no entanglement needed, (2) quantum
memory is no longer required, and (3) less information leaked. Therefore,
better security and great feasibility are both achieved.

\bigskip

\textit{keywords:} Quantum cryptography; Quantum private comparison; Bell states.
\end{abstract}

%\pacs{03.67.Dd, 03.67.Ac}

%\PACS{{03.67.Dd}{Quantum cryptography and communication security} \and
%{03.67.Ac}{Quantum algorithms, protocols, and simulations}
%     } % end of PACS codes

\maketitle

%%%%%%%%%%%%%%%%%%%%%%%%%%%%%%%%%%%%%%%%%%%%%%%%%%%%%%%%%%%%%%%%%%%%%%%%%%%%%%%%%%%%%%%%%%%%%%%%%%%%%%%%%%%%%%%%%%%%%%%%%%%%%%%%%%%%%%%%%%%%%%%%%%%%%%%%%%%%%%%%%%%%%%%%%%%%%%%%%%%%%%%%%%%%%%%%%%%%%%%%%%%%%%%%%%%%%%%%%%%%%%%%%%%%%%%%%%%%%%%%%%%%%%%%%%%%

\section{Introduction}

Private comparison \cite{m2} is a two-party cryptographic problem where
Alice has a private data $a$\ and Bob has a private data $b$. They want to
determine whether $a$ and $b$ are equal, without revealing any extra
information on their values other than what can be inferred from the
comparison result.

It is well-known that unconditionally secure quantum two-party secure
computations are impossible \cite{qi149,qi500,qi499,qi677,qi725,qbc61}.
Therefore, most existing quantum private comparison (QPC) protocols added a
third party to accomplish the task (see Ref. \cite{HeIJQI13} and the references
therein). In 2016, we proposed a QPC protocol \cite{HeIJQI16} which involves
two parties only. Although it is not unconditionally secure, the loose upper
bound of the average amount of information leaked is $14$ bits only. It is
also very feasible because quantum memory and entanglement are not required.
Later, we further proposed the device-independent version of the protocol
\cite{HePS18}.

Recently, Wu, Cai, Wu and Zhang \cite{qi1696} reported that they found our
protocol in Ref. \cite{HeIJQI16} contains two problems: (1) it is insecure
against external eavesdropping, and (2) it is not suitable for comparing
short strings. They also proposed a new one (called as the WCWZ protocol
thereafter) aiming to fix these problems. We feel that the criticism is
improper. But Ref. \cite{qi1696} was not submitted as a Comment paper, so we
did not aware of it until it was published, and we did not have a chance to
make a formal reply. Also, Ref. \cite{qi1696} has not provided a rigorous
calculation on the amount of information leaked in their own protocol as we
did in Ref. \cite{HeIJQI16}, making it hard to judge whether their protocol is
more suitable for comparing short strings.

In the current paper, we will reply to the criticism, and study the amount
of information leaked in the WCWZ protocol so that the performance of the
protocols can be compared clearly. Moreover, we will also propose an
improved protocol, which not only has all the advantages of the WCWZ
protocol (e.g., secure against external eavesdropping, and low amount of
information leaked without the need of a third party), but also requires
much less quantum resources (e.g., quantum memory and entanglement) so that
it becomes much more feasible.

The paper is organized as follows. In the next section, we will review our
previous protocol in Ref. \cite{HeIJQI16}. Then in section III, we will present the
criticism made by Ref. \cite{qi1696} and our reply. The WCWZ protocol \cite{qi1696} will be
reviewed in section IV, and we will analyze its amount of information leaked
and related problems in section V. In section VI, we will propose our
improved protocol. Then we will prove its security in section VII and compare
it with the protocols in Refs. \cite{qi1696} and \cite{HeIJQI16} in section VIII.

\section{Our previous protocol}

Let $H(x)$\ be a classical hash function which is a $1$-to-$1$\ mapping
between the $n$-bit strings $x$ and $y=H(x)$ (i.e., $H:\{0,1\}^{n}%
\rightarrow \{0,1\}^{n}$). Denote the two orthogonal states of a qubit as $%
\left\vert 0\right\rangle _{0}$\ and $\left\vert 1\right\rangle _{0}$,
respectively, and define $\left\vert 0\right\rangle _{1}\equiv (\left\vert
0\right\rangle _{0}+\left\vert 1\right\rangle _{0})/\sqrt{2}$, $\left\vert
1\right\rangle _{1}\equiv (\left\vert 0\right\rangle _{0}-\left\vert
1\right\rangle _{0})/\sqrt{2}$. That is, the subscript $\sigma =0,1$ in $%
\left\vert \gamma \right\rangle _{\sigma }$\ stands for two incompatible
measurement bases, while\ $\gamma =0,1$\ distinguishes the two states in the
same basis. The two-party QPC protocol that we proposed in Ref. \cite{HeIJQI16}
is as follows.

\bigskip

\textit{Our previous QPC Protocol} (for comparing Alice's $n$-bit string $%
a\equiv a_{1}a_{2}...a_{n}$\ and Bob's $n$-bit string $b\equiv
b_{1}b_{2}...b_{n}$):

(1) Using the hash function $H(x)$, Alice calculates the $n$-bit string $%
h^{A}\equiv h_{1}^{A}h_{2}^{A}...h_{n}^{A}=H(a)$, and Bob calculates the $n$%
-bit string $h^{B}\equiv h_{1}^{B}h_{2}^{B}...h_{n}^{B}=H(b)$.

(2) From $i=1$ to $n$, Alice and Bob compare $h^{A}$ and $h^{B}$ bit-by-bit
as follows.

\qquad If $i$ is odd, then:

\qquad \qquad (2.1A) Alice randomly picks a bit $\gamma _{i}^{A}\in \{0,1\}$%
\ and sends Bob a qubit in the state $\left\vert \gamma
_{i}^{A}\right\rangle _{h_{i}^{A}}$.

\qquad \qquad (2.2A) Bob measures it in the $h_{i}^{B}$\ basis and obtains
the result $\left\vert \gamma _{i}^{B}\right\rangle _{h_{i}^{B}}$. He
announces $\gamma _{i}^{B}$\ while keeping $h_{i}^{B}$\ secret.

\qquad \qquad (2.3A) Alice announces $\gamma _{i}^{A}$.

\qquad If $i$ is even, then:

\qquad \qquad (2.1B) Bob randomly picks a bit $\gamma _{i}^{B}\in \{0,1\}$\
and sends Alice a qubit in the state $\left\vert \gamma
_{i}^{B}\right\rangle _{h_{i}^{B}}$.

\qquad \qquad (2.2B) Alice measures it in the $h_{i}^{A}$\ basis and obtains
the result $\left\vert \gamma _{i}^{A}\right\rangle _{h_{i}^{A}}$. She
announces $\gamma _{i}^{A}$\ while keeping $h_{i}^{A}$\ secret.

\qquad \qquad (2.3B) Bob announces $\gamma _{i}^{B}$.

\qquad (2.4) If $\gamma _{i}^{A}\neq \gamma _{i}^{B}$, then they conclude
that $a\neq b$, and abort the protocol immediately without comparing the
rest bits of $h^{A}$ and $h^{B}$. Otherwise they continue with the next $i$.

(3) If Alice and Bob find $\gamma _{i}^{A}=\gamma _{i}^{B}$ for all $%
i=1,...,n$ then they conclude that $a=b$.

\bigskip

As shown by Eqs. (7) and (8) of Ref. \cite{HeIJQI16}, the loose upper bound of
the average amount of mutual information leaked in this protocol is%
%\begin{eqnarray}
%I_{A} &=&\sum_{k=1}^{[n/2]}m\times p_{abort}^{m}  \nonumber \\
%&=&\sum_{k=1}^{[n/2]}2k\cos ^{2(k-1)}(\pi /8)\sin ^{2}(\pi /8)  \label{IA}
%\end{eqnarray}%
\begin{eqnarray}
I_{A}=\sum_{k=1}^{[n/2]}2k\cos ^{2(k-1)}(\pi /8)\sin ^{2}(\pi /8)  \label{IA}
\end{eqnarray}%
for dishonest Alice, and%
%\begin{eqnarray}
%I_{B} &=&\sum_{k=1}^{[(n+1)/2]}m\times p_{abort}^{m}  \nonumber \\
%&=&\sum_{k=1}^{[(n+1)/2]}(2k-1)\cos ^{2(k-1)}(\pi /8)\sin ^{2}(\pi /8)
%\label{IB}
%\end{eqnarray}%
\begin{eqnarray}
I_{B}=\sum_{k=1}^{[(n+1)/2]}(2k-1)\cos ^{2(k-1)}(\pi /8)\sin ^{2}(\pi /8)
\label{IB}
\end{eqnarray}%
for dishonest Bob, where $[x]$ denotes the integer part of $x$.

The purpose of using the hash function in the protocol is to change the
direct information leaked into mutual information. For example, suppose that
$n=3$ and Bob's secret string is $b=011$. Consider that he compares $b$
with Alice's secret string $a$ bit-by-bit directly in the above protocol
without using the hash function. If the protocol aborts at the $i=1$ round,
then Bob knows immediately that the first bit of $a$ must be different
from that of $b$, i.e., $a_{1}=\bar{b}_{1}=1$. Thus, Bob knows that the
possible choices of the value of $a$ must be limited to the set $%
\{100,101,110,111\}$. Note that before running the protocol, from Bob's
view, all the $2^{3}=8$ possible values of an arbitrary $3$-bit string may
be taken by $a$. Therefore, the amount of information leaked to Bob in the
protocol is $\log _{2}8-\log _{2}4=1$ bit. On the contrary, consider that
they use the hash function $H(x)$\ shown in Fig.1 and run the above protocol
faithfully, i.e., they compare $h^{A}=H(a)$\ and $h^{B}=H(b)$\ instead of $a$
and $b$ directly. If the protocol also aborts at the $i=1$ round, since $%
H(b)=H(011)=010$, Bob knows that there must be $h_{1}^{A}=\bar{h}_{1}^{B}=1$%
, so that $H(a)$ must be limited to $\{100,101,110,111\}$. Then from Fig.1
he knows that $a$ must be limited to $\{000,110,101,010\}$. Again, the
amount of information leaked is also $1$ bit. But in this case, we can see
that Bob can no longer be sure whether the first bit of $a$ is $0$ or $1$.
That is, when using the hash function, while the amount of information
leaked remains the same, the type of this information is changed from the
direct information of the secret string of the other party into the mutual
information.

\begin{figure}[t]
\centering
\includegraphics[scale=1.1]{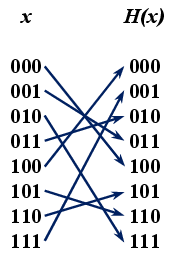}
\caption{An example of the hash function $H(x)$ used in the protocols.}
\label{fig:epsart}
\end{figure}

This could be useful in practical applications. For instance, suppose that
two companies want to compare whether their bids $a$ and $b$\ to a project
is equal or not, while they do not want to reveal whose bid is higher. Without
using the hash function, after running the protocol the first bit will be
leaked to the other party, so that they both know who had placed a higher
bid. But when the hash function is used, generally they merely learn that
their bids are different, without revealing which is the higher one.

\section{WCWZ's criticism and our reply}

In Ref. \cite{qi1696}, Wu, Cai, Wu and Zhang claimed that our above protocol
has two security loopholes. Namely, in their own words, (1)
\textquotedblleft outside party (or called Eve) can obtain Alice's and Bob's
partial private information before aborting the protocol without being
detected\textquotedblright , and (2) \textquotedblleft He's protocol cannot
ensure fairness perfectly ... this cheating strategy leads to the average
amount of mutual information leaked as $13$ bits at most in Ref. \cite%
{HeIJQI16}. So He's original protocol is not suitable for a smaller
bit-length comparison protocol\textquotedblright .

We do not agree with these criticism. First, in literature two-party
cryptographic protocols (including QPC) are required to be secure against
internal cheatings only, while the security against external attacks are not
mandatory. Thus it is improper to call it a security loophole if the attack
does not come from internal legitimate participants, i.e., Alice and Bob.
Second, the $13$ bits of mutual information leaked is merely the loose
upper bound for very long strings. When comparing strings with a smaller
bit-length, the amount of mutual information leaked will drop significantly. Now let us elaborate these two points in details.

(1) As Kilian pointed out \cite{qi139}, \textquotedblleft the reason
two-party protocol problems are so difficult is due to a simple symmetry
condition on what players know about each others data\textquotedblright .
Therefore, most previous studies on two-party cryptography merely interested
in the security against internal cheating from legitimate participants
(i.e., Alice and Bob). The security against the attack from external
eavesdroppers is not considered as an obligated task of two-party secure
computation protocols. For example, in various no-go proofs of
unconditionally secure quantum bit commitment \cite%
{qi24,qi23,qi58,qi581,qi611,qi147,qbc49,qbc35,qi610,qi283,qi323,qi714,qbc12,qbc25,qbc32,qi715,qbc182}
and cheat sensitive bit commitment \cite{qi150,qi442,HeSR}, coin flipping
\cite{qi58,qi76,qi817,qi145,qbc37,qi246,qbc132,qbc19}, quantum seals \cite%
{impossibility,HeSeal}, oblivious transfer \cite{*qbc40,qbc14} and two-party
secure computations \cite{qi149,qi500,qi499,qi677,qi725,qbc61}, the security
of the protocols is always discussed with internal attacks only. Especially,
there were proposals on relativistic bit commitment \cite%
{qi44,qi582,qbc24,qbc51,qbc69,qbc66,qbc83,qbc82,qbc105},
spacetime-constrained oblivious transfer \cite{qbc130,qbc201,qbc216} and
quantum token \cite{qbc224} which are all accepted as unconditionally secure,
but the security proofs against external attacks are not presented either.

That is, in literature two-party secure computation protocols were regarded
as secure as long as it can defeat internal attacks. It is not considered as
a \textquotedblleft loophole\textquotedblright\ even if the external party
Eve can cheat in these protocols. This is also the case of our previous QPC
protocol in Ref. \cite{HeIJQI16}.

(2) In section 3.2 of the WCWZ paper \cite{qi1696}, the authors claimed that
\textquotedblleft this subsection points that He's protocol cannot ensure
fairness perfectly. The result of comparison can be manipulated partially by
either party\textquotedblright . But as we were fully aware that
unconditionally secure quantum two-party secure computations are considered
impossible \cite{qi149,qi500,qi499,qi677,qi725,qbc61}, we already presented
clearly in the abstract of Ref. \cite{HeIJQI16} that our purpose was to
\textquotedblleft study how far we can go with two parties
only\textquotedblright , and in our protocol \textquotedblleft the average
amount of information leaked cannot be made arbitrarily
small\textquotedblright . We were surprised that the WCWZ paper made it
sound like as if they found a new loophole that we missed to mention.

The second paragraph of section 3.2 of Ref. \cite{qi1696} went on to stated
that \textquotedblleft the cheating strategy is simple ... this cheating
strategy leads to the average amount of mutual information leaked as $13$
bits at most in Ref. \cite{HeIJQI16}. So He's original protocol is not
suitable for a smaller bit-length comparison protocol\textquotedblright .
However, their cheating strategy\ is exactly what we already described in
the security analysis in section 4 of Ref. \cite{HeIJQI16}. We also clearly presented that the average
amount of mutual information leaked is below $13$\ bits for dishonest Bob ($%
14$\ bits for dishonest Alice), which was obtained from Eqs. (7) and (8) of
Ref. \cite{HeIJQI16} (i.e., Eqs. (\ref{IA}) and (\ref{IB}) of the current paper).
Thus this is not a new result either.
But as we also elaborated in section 4 of Ref. \cite{HeIJQI16}, this amount of mutual
information leaked is merely the \textit{loose} upperbound so that it can be
\textquotedblleft sufficiently general to cover any kind of cheating
strategies potentially existed\textquotedblright . Any currently known cheating
strategy, including the one that section 3.2 of Ref. \cite{qi1696} cited from us, cannot actually saturate this bound. In fact, as we shown in the last paragraph of page 6 of Ref. \cite{HeIJQI16}, when using the cheating strategy that Ref. \cite{qi1696} cited, the cheater can optimally discriminate $\gamma _i$ (in Ref. \cite{qi1696} it was written as $x_i$) but has absolutely zero knowledge on the value of the hash bit $h_i$ of the other party in half of the rounds. Therefore, in Eqs. (\ref{IA}) and (\ref{IB}) the first terms \textquotedblleft $2k$\textquotedblright\ and \textquotedblleft $2k-1$\textquotedblright\ will be replaced by \textquotedblleft $k$\textquotedblright\ and \textquotedblleft $k-1$\textquotedblright , respectively. Then the bound for this specific cheating strategy will drop to $7$ bits for dishonest Alice and $6$ bits for dishonest Bob. More importantly, the general bound \textquotedblleft $%
13 $\ bit\textquotedblright\ is for dishonest Bob in the limit $%
n\rightarrow \infty $ only, where $n$\ denotes the length of the bit-strings
being compared. When $n<60$\ bit, the average amount of mutual information
leaked will be significantly reduced, as shown in Fig.1 of Ref. \cite{HeIJQI16} (see also point (3) of section V.C of the current paper).

Meanwhile, Ref. \cite{qi1696} has not provided a rigorous calculation on the
amount of information leaked in their own protocol. In the next section, we
will review the WCWZ protocol. Then a rigorous calculation on its amount of
information leaked and comparison with our protocol will be provided at the
end of section V.C and in section VIII.A, where we will see that our protocol
is actually more suitable for comparing short strings than their protocol
does.

\section{The WCWZ protocol}

In section 3.3 of Ref. \cite{qi1696}, Wu, Cai, Wu and Zhang proposed the
following protocol.

\bigskip

\textit{The WCWZ Protocol:}

Step 1. Using a 1-to-1 classical hash function $H:\{0,1\}^{n}\rightarrow
\{0,1\}^{n}$, Alice computes the $n$-bit string $H(a)=h_{1}^{A}...h_{n}^{A}$%
\ of secret information $a$, and Bob computes the $n$-bit string $%
H(b)=h_{1}^{B}...h_{n}^{B}$\ of secret information $b$.

Step 2. Alice divides the value $H(a)$\ into $\left\lceil n/m\right\rceil $\
($m\geq 2$) groups, which are%
\begin{eqnarray}
X_{0} &=&\{h_{1}^{A},...,h_{m}^{A}\}  \nonumber \\
X_{1} &=&\{h_{m+1}^{A},...,h_{2m}^{A}\}  \nonumber \\
&&...  \nonumber \\
X_{\left\lceil \frac{n}{m}\right\rceil -1} &=&\{h_{\left\lceil \frac{n}{m}%
\right\rceil \ast m+1}^{A},...,h_{n-1}^{A}\}.
\end{eqnarray}%
Bob does the same operation as Alice and obtains%
\begin{eqnarray}
Y_{0} &=&\{h_{1}^{B},...,h_{m}^{B}\}  \nonumber \\
Y_{1} &=&\{h_{m+1}^{B},...,h_{2m}^{B}\}  \nonumber \\
&&...  \nonumber \\
Y_{\left\lceil \frac{n}{m}\right\rceil -1} &=&\{h_{\left\lceil \frac{n}{m}%
\right\rceil \ast m+1}^{B},...,h_{n-1}^{B}\}.
\end{eqnarray}%
(\textit{While we believe that the last terms }$h_{n-1}^{A}$\textit{\ and }$%
h_{n-1}^{B}$\textit{\ in these two equations\ should be }$h_{n}^{A}$\textit{%
\ and }$h_{n}^{B}$\textit{, respectively, here we present the original form
of the protocol in Ref. \cite{qi1696} as is.})

Step 3. Alice (Bob) prepares $m$ Bell states as initial states, every Bell
state is randomly chosen from $\left\vert \Phi ^{+}\right\rangle
=(\left\vert 00\right\rangle +\left\vert 11\right\rangle )/\sqrt{2}$, $%
\left\vert \Psi ^{+}\right\rangle =(\left\vert 01\right\rangle +\left\vert
10\right\rangle )/\sqrt{2}$. Alice (Bob) records these initial states as $%
S_{A}$\ ($S_{B}$). The first particles of all Bell states $S_{A}$\ ($S_{B}$%
)\ form the sequence $S_{A_{1}}$\ ($S_{B_{1}}$), and the rest form the
sequence $S_{A_{2}}$\ ($S_{B_{2}}$).

%\textit{(For example, suppose that Alice prepares the particles }$P_{a}$%
%\textit{, }$P_{b}$\textit{, }$P_{c}$\textit{, }$P_{d}$\textit{, }$P_{e}$%
%\textit{, }$P_{f}$\textit{, ... in the states }$\left\vert
%P_{a}P_{b}\right\rangle =\left\vert \Phi ^{+}\right\rangle $\textit{, }$%
%\left\vert P_{c}P_{d}\right\rangle =\left\vert \Psi ^{+}\right\rangle $%
%\textit{, }$\left\vert P_{e}P_{f}\right\rangle =\left\vert \Phi
%^{+}\right\rangle $\textit{, Then }$S_{A}\equiv \left\vert
%P_{a}P_{b}\right\rangle \otimes \left\vert P_{c}P_{d}\right\rangle \otimes
%\left\vert P_{e}P_{f}\right\rangle \otimes ...$\textit{, }$S_{A_{1}}\equiv
%P_{a}P_{c}P_{e}...$\textit{, and }$S_{A_{2}}\equiv P_{b}P_{d}P_{f}...$%
%\textit{\ . Bob's }$S_{B}$\textit{, }$S_{B_{1}}$\textit{, and }$S_{B_{2}}$%
%\textit{\ are defined in a similar way.)}

Step 4. Alice (Bob) prepares decoy states $D_{A}$\ ($D_{B}$), randomly in
states $\left\vert 0\right\rangle $, $\left\vert 1\right\rangle $, $%
(\left\vert 0\right\rangle +\left\vert 1\right\rangle )/\sqrt{2}$, $%
(\left\vert 0\right\rangle -\left\vert 1\right\rangle )/\sqrt{2}$. Alice
(Bob) randomly inserts $D_{A}$\ ($D_{B}$) in $S_{A_{1}}$\ ($S_{B_{1}}$) to
form a new sequence $S_{A_{1}}^{\prime }$\ ($S_{B_{1}}^{\prime }$), then
sends it to Bob (Alice).

%\textit{(For example, Alice can prepare the decoy states as }$D_{A}\equiv
%\left\vert D_{1}\right\rangle \otimes \left\vert D_{2}\right\rangle \otimes
%\left\vert D_{3}\right\rangle \otimes \left\vert D_{4}\right\rangle \otimes
%...$\textit{\ where }$\left\vert D_{1}\right\rangle =\left\vert
%0\right\rangle $\textit{, }$\left\vert D_{2}\right\rangle =(\left\vert
%0\right\rangle +\left\vert 1\right\rangle )/\sqrt{2}$\textit{, }$\left\vert
%D_{3}\right\rangle =(\left\vert 0\right\rangle -\left\vert 1\right\rangle )/%
%\sqrt{2}$\textit{, }$\left\vert D_{4}\right\rangle =\left\vert
%1\right\rangle $\textit{, ... . Then a possible choice for }$%
%S_{A_{1}}^{\prime }$\textit{\ can be }$S_{A_{1}}^{\prime }\equiv
%D_{1}D_{2}P_{a}D_{3}D_{4}D_{5}P_{c}D_{6}P_{e}D_{7}D_{8}...$\textit{\ .)}

Step 5. After confirming that Bob (Alice) has received the quantum sequence $%
S_{A_{1}}^{\prime }$\ ($S_{B_{1}}^{\prime }$), Alice (Bob) informs the
positions and the measurement bases of $D_{A}$\ ($D_{B}$) to Bob (Alice).
Subsequently, Bob (Alice) extracts the particles in $D_{A}$\ ($D_{B}$) from $%
S_{A_{1}}^{\prime }$\ ($S_{B_{1}}^{\prime }$), and gets the sequences $%
S_{A_{1}}$\ ($S_{B_{1}}$). Therefore, Alice and Bob can check the existence
of an Eve by a predetermined threshold of error rate. If the error rate is
limited to the predetermined threshold, there is no Eve and the protocol
continues. Otherwise, Alice and Bob abort the protocol and restart from step
1.

Step 6. Bob (Alice), respectively, performs $X=\left\vert 1\right\rangle
\left\langle 0\right\vert +$\ $\left\vert 0\right\rangle \left\langle
1\right\vert $ or $I=\left\vert 0\right\rangle \left\langle 0\right\vert +$\
$\left\vert 1\right\rangle \left\langle 1\right\vert $ operation on the $i$%
th particle of sequence $S_{A_{1}}$\ ($S_{B_{1}}$) when $h_{i}^{B}=1$\ ($%
h_{i}^{A}=1$) or $h_{i}^{B}=0$\ ($h_{i}^{A}=0$), and obtains the sequence $%
S_{A_{1}}^{\prime \prime }$\ ($S_{B_{1}}^{\prime \prime }$). Then, Bob
(Alice) randomly inserts $D_{A}^{\prime \prime }$\ ($D_{B}^{\prime \prime }$%
) in $S_{A_{1}}^{\prime \prime }$\ ($S_{B_{1}}^{\prime \prime }$) and forms
a new sequence $S_{A_{1}}^{\prime \prime \prime }$\ ($S_{B_{1}}^{\prime
\prime \prime }$).

Step 7. Bob sends sequence $S_{A_{1}}^{\prime \prime \prime }$\ to Alice,
and Alice sends the sequence $S_{B_{1}}^{\prime \prime \prime }$\ to Bob.
After confirming receipt of the sequences $S_{A_{1}}^{\prime \prime \prime }$%
\ and $S_{B_{1}}^{\prime \prime \prime }$, Alice and Bob publish the
positions and the measurement bases of $D_{A}^{\prime }$\ and $D_{B}^{\prime
}$ (\textit{we believe that the authors meant the decoy states }$%
D_{A}^{\prime \prime }$\textit{\ and }$D_{B}^{\prime \prime }$)
simultaneously. Alice and Bob recover the sequences $S_{A_{1}}^{\prime
\prime }$\ and $S_{B_{1}}^{\prime \prime }$ through discarding the decoy
states, individually. Alice and Bob check for the existence of an Eve, as
described in step 5. If there is no Eve and the protocol continues.
Otherwise, Alice and Bob abort the protocol and restart from step 1.

Step 8. Bob (Alice) performs $X=\left\vert 1\right\rangle \left\langle
0\right\vert +$\ $\left\vert 0\right\rangle \left\langle 1\right\vert $ or $%
I=\left\vert 0\right\rangle \left\langle 0\right\vert +$\ $\left\vert
1\right\rangle \left\langle 1\right\vert $ operation on the $i$th particle
of sequence $S_{A_{1}}^{\prime \prime }$\ ($S_{B_{1}}^{\prime \prime }$)
when $h_{i}^{B}=1$\ ($h_{i}^{A}=1$) or $h_{i}^{B}=0$\ ($h_{i}^{A}=0$), and
obtains the new state $S_{B}^{\prime }$\ ($S_{A}^{\prime }$). If $%
S_{B}^{\prime }=S_{B}$\ ($S_{A}^{\prime }=S_{A}$), then Bob (Alice)
announces that the compared secret information are identical after
measurements. Otherwise, Bob (Alice) announces the comparison which are
regarded as different.

\section{Cryptanalysis of the WCWZ protocol}

\subsection{A trivial problem}

This protocol contains an obvious problem, probably came from typos. That
is, the secret information $a$ and $b$ that Alice and Bob want to compare
are $n$-bit strings, but in step 3 they exchange $m$ Bell states only (where
$m<n$, as can be seen from step 2), i.e., only the first $m$ bits of the
hash values $H(a)$ and $H(b)$ are compared. Consequently, there will be the
problem that if $H(a)\neq H(b)$ while the first $m$ bits of $H(a)$ and $H(b)$
happen to be identical, their protocol will mistakenly output $H(a)=H(b)$ in
step 8 as the final result.

There is a trivial fix to this problem. In step 8 when Alice and Bob found
the $m$ bits being compared are identical, they should repeat steps 3-8
again, exchanging another set of $m$ Bell states to compare the next $m$
bits of $H(a)$ and $H(b)$. They repeat this procedure over and over until
there is a run in which step 8 shows that some bits of $H(a)$ and $H(b)$\
are different, or until all bits are compared and shown to be identical.
With this modification, the protocol will always output the correct result.
Actually we believe that this is exactly what the authors had in mind. But
without this modification explicitly written, the original protocol in Ref. \cite{qi1696} cannot be regarded as correct.

\subsection{Simultaneity problem}

In step 7 of the WCWZ protocol, Alice and Bob are required to
\textquotedblleft publish the positions and the measurement bases of $%
D_{A}^{\prime }$\ and $D_{B}^{\prime }$ \textit{simultaneously}%
\textquotedblright . But such a requirement is generally considered
inappropriate in nonrelativistic cryptography. This is because it is
well-accepted \cite{qi134} that \textquotedblleft the standard
nonrelativistic cryptographic scenario for two mistrustful parties is as
follows ... In particular, neither of them has any way of ensuring that a
message sent by the other was sent a certain time before receipt, and so an
effectively simultaneous exchange of messages cannot be arranged. A standard
cryptographic protocol thus prescribes a sequential exchange of messages
between A and B, in which message $i+1$\ is not sent until the sender has
received message $i$.\textquotedblright\ Indeed, if the security of a
nonrelativistic protocol relies on simultaneity, then it leaves room for
potential cheats. For example, consider that the distance between Alice's
and Bob's sites is $L$, and they are supposed to receive messages from each
other at time $t_{1}$ which is measured in the same reference frame
stationary to both of them. If they communicate with methods in which the
message carrier travels with the speed of light $c$, then each of them should
start sending her/his message at time $t_{0}\equiv t_{1}-L/c$. But dishonest
Bob may set up an \textquotedblleft agent\textquotedblright\ site secretly,
which is $L/3$ away from Alice. Then if honest Alice sends her message at
time $t_{0}$, Bob's agent site will receive it at time $t_{1}^{\prime
}\equiv t_{0}+(1/3)L/c<t_{1}$. In this case, Bob has a time interval with
the duration $(1/3)L/c$\ to analysis the message received from Alice, and
decide the content of the message to be sent which could benefit his
cheating, and send it at time $t_{1}^{\prime \prime }\equiv t_{0}+(2/3)L/c$.
His message will still reach Alice at time $t_{1}$ so that this cheating
cannot be detected. But he manages to delay the sending of his message until
he receives Alice's message so that the simultaneity is broken.

%\textquotedblleft Relativity plays an essential role in this
%protocol: its security is guaranteed by the impossibility of superluminal
%signaling.\textquotedblright\ \cite{qi44}

On the contrary, in relativistic cryptography there is countermeasure
against the above cheating, so that simultaneous exchange of messages can be
available. Refs. \cite{qi44,qi582,qi134} described such an arrangement. Alice and
Bob agree on a frame and two locations $x_{1}$, $x_{2}$. Honest Alice (Bob)
is supposed to erect her (his) laboratory near $x_{1}$ ($x_{2}$), within an
agreed distance $\delta \ll \left\vert x_{1}-x_{2}\right\vert $. To verify
that Alice is indeed there, in relativistic cryptography Bob is allowed to
have an \textquotedblleft agent\textquotedblright\ $B_{1}$ near $x_{1}$ \cite{qi582}. At
any time $B_{1}$ can send a test signal to Alice and expect to receive a
response within time $2\delta /c$, so that he can confirm to Bob about
Alice's location. Similarly, Alice can test whether Bob is indeed near $%
x_{2} $ via her own agent. Given that superluminal signaling is impossible,
in this setup once Alice sends a message $m_{A}$ to Bob and then receives
Bob's message $m_{B}$ within time $t\ll 2(\left\vert x_{1}-x_{2}\right\vert
-2\delta )/c$, she can be sure that Bob had sent $m_{B}$ before he received $%
m_{A}$. The same criterion also holds for Bob. In this case, we can take $%
m_{A}$ and $m_{B}$ as messages being exchanged simultaneously even if their
actual sending times may differ slightly, because it makes no difference to
the security analysis.

Therefore, if a protocol assumes that there is an approach to force both
Alice and Bob to send messages simultaneously, then it is actually a
relativistic protocol. In this scenario, it is not surprising that hard
cryptographic tasks will become much easier. For example, coin flipping (CF)
can be easily realized even without quantum methods, as long as simultaneity
becomes available \cite{qi134} (please see the appendix for a brief review
on the protocol). On the other hand, it is a widely-accepted result that nonrelativistic
unconditionally secure CF with an arbitrarily small bias is impossible \cite%
{qi58,qi76,qi817,qi145,qbc37,qi246,qbc132,qbc19}. This is another evidence
showing that in literature, simultaneity is not accepted in nonrelativistic
cryptography.

For this reason, it is unfair to compare the WCWZ protocol (which relies on
simultaneity) with our nonrelativistic protocol in Ref. \cite{HeIJQI16} (where
simultaneity is not needed). In section VI, we will propose another protocol
in which simultaneity is also assumed to be available, so that it can be
compared with the WCWZ protocol in the same scenario.

\subsection{Information leaked}

In Ref. \cite{qi1696} the authors did not give a rigorous evaluation on the
amount of information leaked in their protocol. Here we provide such an
evaluation.

Since Alice and Bob compare $m$ bits of $H(a)$ and $H(b)$ all at the same
time, they will always know $m$ bits of the other's data no matter the
comparison result is identical or not. Therefore, even when nobody cheats,
the protocol leaks at least $m$ bits of information to the other party, with
or without including the modification in section V.A.

Now consider the case where the above modification is included, i.e., if the
first $m$ bits of $H(a)$ and $H(b)$ are found to be identical, Alice and Bob
continue to compare the rest bits. Given that the hash function $H(x)$\ ($%
x\in \{a,b\}$) is a random\ mapping between $x$ and $y=H(x)$, each pair of
the hash bits $h_{i}^{A}$ and $h_{i}^{B}$\ stands probability $1/2$ to be
identical. Consequently, when Alice and Bob compare the first $m$ bits of $%
H(a)$ and $H(b)$ using the WCWZ protocol, the probability for finding all
these $m$ bits to be identical in step 8 (so that the protocol continues) is%
\begin{equation}
p_{m}=\left( \frac{1}{2}\right) ^{m},
\end{equation}%
and the probability for finding at least one of these $m$ bits to be
different in step 8 (so that the protocol aborts) is then%
\begin{equation}
p_{a}=1-p_{m}=1-\frac{1}{2^{m}}.
\end{equation}

If the protocol indeed aborts, then both Alice and Bob know these $m$ bits
of mutual information on the other's private data. That is, there is
probability $p_{a}$ that the amount of information leaked is $m$ bits.

Else if these $m$ bits are identical (which occurs with probability $p_{m}$)
and Alice and Bob continue to compare the next $m$ bits by repeating steps
3-8 for the $2$nd round, then again there will be probability $p_{a}$ that
the next $m$ bits are different so that the protocol aborts after the $2$nd
round. In this case both parties know the first $2m$ bits. That is, with
probability $p_{m}p_{a}$\ the protocol will abort at the $2$nd round, and
the amount of information leaked is $2m$ bits.

Continue with this analysis, we can see that the probability for the
protocol to abort at the $i$-th round is%
\begin{equation}
p_{i}=(p_{m})^{i-1}p_{a}=\frac{1}{2^{m(i-1)}}\left( 1-\frac{1}{2^{m}}\right)
,
\end{equation}%
and the amount of information leaked is $im$ bits.

For simplicity, suppose that $n/m$ is an integer. Then summing over all
possible $i$ values, we finally yield the average amount of mutual
information leaked%
\begin{equation}
I=\sum_{i=1}^{\left\lceil n/m\right\rceil }(im\times
p_{i})=m\sum_{i=1}^{\left\lceil n/m\right\rceil }\frac{i(2^{m}-1)}{2^{im}}.
\label{I}
\end{equation}%
Note that this value does not included the amount of information leaked when
the protocol never aborts in the middle, but continues until all bits are
compared and found to be identical instead. This is correct, because when $a$
and $b$ are identical, both Alice and Bob surely know all the $n$ bits, and
this is allowed since by definition a QPC protocol is secure as long as it
does not reveal any extra information on the compared values
\textquotedblleft other than what can be inferred from the comparison
result\textquotedblright .

\begin{figure}[htbp]
\centering
\includegraphics[scale=0.85]{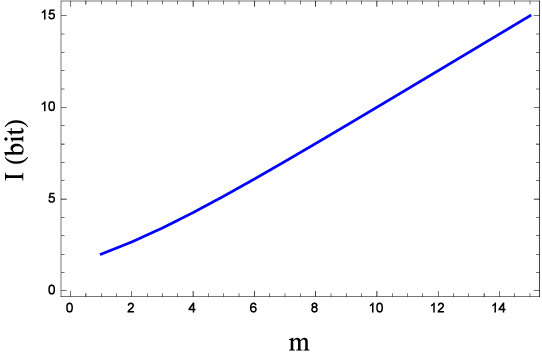}
\caption{The average amount of information leaked $I$ as a function of the parameter $m$ in the WCWZ protocol when the length of the strings being compared is $n=360360$.}
\label{fig:epsart}
\end{figure}

Fig.2 shows $I$ as a function of the value of $m$, as calculated from Eq. (%
\ref{I}), with the length of the compared strings $a$ and $b$ is fixed as $%
n=360360$ (so that $n/m$ is always an integer for $m=2,...,15$). From the figure we can find the following results:

(1) The value of $I$\ grows as $m$ increases. It indicates that introducing $%
m$ in the WCWZ protocol is completely unnecessary, and taking $m\geq 2$\ in
step 2 is not a wise choice. The smaller the value of $m$ is, the less
amount of information will be leaked to Alice and Bob. Thus the optimal
choice is $m=1$. That is, Alice and Bob should not divide $H(a)$ and $H(b)$
into $\left\lceil n/m\right\rceil $ groups in step 2. Instead, they should
better compare them bit-by-bit, like we did in our protocol in Ref. \cite%
{HeIJQI16}.

(2) If Alice and Bob choose $m\geq 14$, then there is $I\geq 14$ bits. In
this case the WCWZ protocol is always less secure than ours in Ref. \cite%
{HeIJQI16} for any length of the strings being compared.

(3) When comparing very short strings, the WCWZ protocol is also less secure
than ours in Ref. \cite{HeIJQI16} even for $m<14$. For example, when $n=6$, Eq. (%
\ref{I}) gives that $I\simeq 2.53$\ bits when $m=2$, and $I\simeq 1.88$\
bits when $m=1$. On the contrary, when using our protocol in Ref. \cite{HeIJQI16}%
, Eqs. (\ref{IA}) and (\ref{IB}) (i.e., Eqs. (7) and (8) of Ref. \cite{HeIJQI16})
show that the average amount of information leaked for $n=6$ is merely $%
I\simeq 1.43$ bits for dishonest Alice and $I\simeq 1.05$ bits for dishonest
Bob, both are smaller than these of the WCWZ protocol.

\subsection{Feasibility problems}

The WCWZ protocol is also very costly in terms of quantum resource.

First, it takes a great amount of quantum memory. As can be seen from steps
5 and 7, whenever Alice and Bob receive the sequences $S_{A_{1}}^{\prime }$,
$S_{B_{1}}^{\prime }$, $S_{A_{1}}^{\prime \prime \prime }$\ and $%
S_{B_{1}}^{\prime \prime \prime }$, they need to wait for the other party to
publish the positions of the decoy states $D_{A}$, $D_{B}$, $D_{A}^{\prime
\prime }$\ and $D_{B}^{\prime \prime }$ before they perform measurements on
them. Otherwise, they may accidently measure the Bell states, while they
should have performed unitary operations $X$ or $I$ on them in steps 6 and 8
instead. Suppose that each of the sequences $D_{A}$, $D_{B}$, $D_{A}^{\prime
\prime }$\ and $D_{B}^{\prime \prime }$ contains $k$ decoy qubits, then the
protocol totally needs the quantum memory for storing $2(k+2m)$ qubits
(i.e., $(k+2m)$ for Alice to store $S_{A_{2}}$ and $S_{B_{1}}^{\prime }$ (or
$S_{A_{1}}^{\prime \prime \prime }$), and the other $(k+2m)$\ for Bob to
store $S_{B_{2}}$ and $S_{A_{1}}^{\prime }$ (or $S_{B_{1}}^{\prime \prime
\prime }$)).

Second, it requires quantum entanglement. Step 3 shows that for each
compared bit, both Alice and Bob need to prepare a pair of Bell state. Thus,
to compare two $n$-bit strings $a$ and $b$, the protocol totally needs $2n$
pairs of Bell states.

To this day, the technology for handling entangled states is still far from
perfect. Long-term storage for quantum states is even more challenging.
Thus, the above requirements make the WCWZ protocol very infeasible.

\section{Our improved protocol}

From the above analysis, we can see that the WCWZ protocol can be improved
in many ways. Here we propose the following one.

\bigskip

\textit{Our Improved Protocol:}

Step i. Alice and Bob perform a quantum key distribution (QKD) protocol
(e.g., the BB84 protocol \cite{qi365}) to share two classical random key strings $%
k^{A}=k_{1}^{A}...k_{n}^{A}$\ and $k^{B}=k_{1}^{B}...k_{n}^{B}$. That is, at
the end of this process, they both know $k^{A}$\ and $k^{B}$\ while the QKD
protocol can keep $k^{A}$\ and $k^{B}$\ secret from any external
eavesdropping.

Step ii. Using a 1-to-1 classical hash function $H:\{0,1\}^{n}\rightarrow
\{0,1\}^{n}$, Alice computes the $n$-bit hash value $%
H(a)=h_{1}^{A}...h_{n}^{A}$\ of her secret information $a$, and Bob computes
the $n$-bit hash value $H(b)=h_{1}^{B}...h_{n}^{B}$\ of his secret
information $b$. Then they compare $H(a)$ and $H(b)$\ bit-by-bit, i.e., for
each single pair of $h_{i}^{A}$ and $h_{i}^{B}$ ($i=1,...n$):

Step iii. Alice announces $c_{i}^{A}=h_{i}^{A}\oplus k_{i}^{A}$ and Bob
announces $c_{i}^{B}=h_{i}^{B}\oplus k_{i}^{B}$ simultaneously.

Step iv. Alice calculates $h_{i}^{B}=c_{i}^{B}\oplus k_{i}^{B}$ and Bob
calculates $h_{i}^{A}=c_{i}^{A}\oplus k_{i}^{A}$.

Step v. Now both Alice and Bob know $h_{i}^{A}$ and $h_{i}^{B}$.

\qquad If they find $h_{i}^{A}=h_{i}^{B}$, they repeat steps iii-iv to
compare the next pair of $h_{i}^{A}$ and $h_{i}^{B}$. If all pairs were
compared and found to be identical, they both know that $a=b$.

\qquad Else if they find $h_{i}^{A}\neq h_{i}^{B}$, they both know that $%
a\neq b$ (but do not announce this comparison result publicly). Unlike
previous protocols, they do not abort the protocol at this stage. Instead,
they replace the rest bits of $H(a)$ and $H(b)$ that have not been compared
(i.e., $h_{i+1}^{A}h_{i+2}^{A}...h_{n}^{A}$ and $%
h_{i+1}^{B}h_{i+2}^{B}...h_{n}^{B}$) with random meaningless bits irrelevant
with $a$, $b$, $H(a)$ and $H(b)$, and repeat step iii to compare these
bits until step iii was totally repeated for $n$ times. The
comparison result of these meaningless bits is not important. The purpose of doing so is merely to puzzle potential eavesdropper Eve, so that she
cannot learn whether $a=b$ or $a\neq b$ by observing whether the protocol
aborts in the middle before all the $n$ bits are compared.

\bigskip

Note that in step iii we assumed that Alice and Bob can announce information
simultaneously, just like they did in step 7 of the WCWZ protocol. If
this simultaneity is not available, the protocol needs a minor modification.
That is, similar to our protocol in Ref. \cite{HeIJQI16}, when $i$ is odd (even),
let Alice announce the information after (before) Bob does. With this
modification, the amount of information leaked to internal cheaters will
become a little higher, because in the odd (even) rounds, dishonest Alice
(Bob) may alter her (his) announcement basing on what the other party
already announced, thus increase the probability for finding $%
h_{i}^{A}=h_{i}^{B}$\ in this round, like it was elaborated in section 4 of
Ref. \cite{HeIJQI16}.

As the WCWZ protocol made use of the existence of simultaneity, for a fair
comparison, in the following security analysis we also take simultaneity as
available, i.e., we only study our protocol with the original form of its
step iii without including the above modification.

\section{Security of our improved protocol}

\subsection{External eavesdropping}

It is easy to prove that an external eavesdropper Eve cannot learn the hash
values $h_{i}^{A}$ and $h_{i}^{B}$, so that she cannot know Alice's and
Bob's secret information $a$ and $b$. This is because $c_{i}^{A}$ and $%
c_{i}^{B}$ were publicly announced in step iii so that Eve surely knows
them. If she can have a strategy to learn either $h_{i}^{A}$ or $h_{i}^{B}$,
then she can learn the secret key $k_{i}^{A}$ or $k_{i}^{B}$ by calculating $%
k_{i}^{A}=c_{i}^{A}\oplus h_{i}^{A}$ or $k_{i}^{B}=c_{i}^{B}\oplus h_{i}^{B}$%
. However, $k_{i}^{A}$ and $k_{i}^{B}$ were generated in step i by the QKD
protocol. it is well known that QKD protocols can be unconditionally secure,
as proven in Ref. \cite{qi70}. Therefore, the existence of any strategy for Eve
to learn $h_{i}^{A}$ or $h_{i}^{B}$\ will conflict with the security proof
of QKD. Thus we know that our protocol is unconditionally secure against
external eavesdropping.

Also, Eve cannot spoil the protocol (i.e., mislead Alice and Bob to a wrong
comparison result). This is also because the QKD protocol ensures that the
key $k_{i}^{A}$ and $k_{i}^{B}$ cannot be altered. Meanwhile, $c_{i}^{A}$
and $c_{i}^{B}$ were publicly announced classically, so that Eve cannot
change them either. Consequently, Alice's and Bob's calculation results of $%
h_{i}^{A}$ and $h_{i}^{B}$\ in step iv will always be correct, leaving no
chance for external eavesdroppers to turn them into the opposite values.

\subsection{Internal attack}

Obviously, when the protocol outputs $h_{i}^{A}\neq h_{i}^{B}$ in
the $i$-th round, both Alice and Bob surely know that the first $i-1$ bits
of $h_{i}^{A}$ and $h_{i}^{B}$\ are identical, while the $i$-th bits are
different. There is no secret in these bits. Therefore, the goal of a
dishonest internal party is to try to make the case $h_{i}^{A}\neq h_{i}^{B}$
occur as late as possible, so that he can learn more bits of data of the
other party. Now we show that this is impossible.

When both parties are honest and the hash function $H(x)$ is a random
mapping between $x$ and $y=H(x)$, the average probability for $%
h_{i}^{A}=h_{i}^{B}$\ (so that the protocol does not abort) is $1/2$. To
show that a dishonest party cannot cheat, we need to show that this
probability cannot be increased.

As the protocol is symmetrical, without loss of generality, let us assume
that Alice is dishonest. Before step iii of the protocol, Bob has not
announced $c_{i}^{B}$, so that Alice does not know the value of $h_{i}^{B}$.
Therefore, by the time that Alice needs to announce $c_{i}^{A}$ in step iii,
she does not know which value can stand a higher probability to make the
calculation of the other party in step iv result in $h_{i}^{A}=h_{i}^{B}$. Also, in step iii we
assumed that Alice and Bob can announce information \textit{simultaneously},
like the WCWZ protocol did. Then by the time that Alice knows Bob's $%
h_{i}^{B}$ from his announced $c_{i}^{B}$, she has also announced $c_{i}^{A}$
to Bob so that she cannot change it anymore.

Consequently, no matter she announces $c_{i}^{A}$ honestly or not, the
probability for $h_{i}^{A}=h_{i}^{B}$\ will still be $1/2$ for any $i$.
%, with or without the cheating from dishonest party.
For the same reason, dishonest
Bob cannot increase the probability for $h_{i}^{A}=h_{i}^{B}$\ either. Then
repeating the reasoning in section V.C, we find that the average amount of
mutual information leaked in our protocol is also described by Eq. (\ref{I}%
), where $m=1$ since $h_{i}^{A}$ and $h_{i}^{B}$\ are compared bit-by-bit in
our protocol. The relationship between $I$ (the average amount of
information leaked) and $n$ (the length of the strings being compared) is
shown as the green line in Fig. 3.

\begin{figure}[b]
\centering
\includegraphics[scale=0.9]{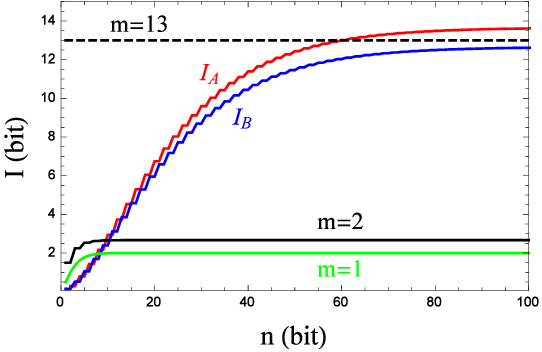}
\caption{The average amount of information leaked $I$ as a function of the length $n$ of the strings being compared. The green line ($m=1$) is for our improved protocol. The black solid (dashed) line is for the WCWZ protocol when $m=2$ ($m=13$). The red (blue) line is the loose upper bound of $I$ for dishonest Alice (Bob) in our previous protocol.}
\label{fig:epsart}
\end{figure}

As we mentioned in the previous section, when simultaneity is not available,
the protocol needs modification. Then the amount of information leaked will
become higher because dishonest party may increase the probability for
finding $h_{i}^{A}=h_{i}^{B}$\ in half of the rounds. But since step 7 of
the WCWZ protocol also makes use of the existence of simultaneity, for a
fair comparison between the protocols, here we assume that simultaneity is
available for our improved protocol too, without taking the above mentioned
modification into account.

\section{Advantages of our improved protocol}

\subsection{Security}

Fig. 3 illustrated the comparison between the three protocols. The green
line represents the performance of our above improved protocol, where the
average amount of mutual information leaked $I$ is calculated from Eq. (\ref%
{I}) by taking $m=1$. The black solid (dashed) line is corresponding to the $%
m=2$ ($m=13$) case of the WCWZ protocol. The red (blue) line indicates the
loose upper bound of the average amount of mutual information leaked to
Alice (Bob) in our previous protocol in Ref. \cite{HeIJQI16}, which is calculated
from Eq. (7) (Eq. (8)) of Ref. \cite{HeIJQI16}. From the comparison we find the
following results.

(I) Since the WCWZ protocol suggested to take $m\geq 2$ in its step 2, it is
always less secure (i.e., the amount of information leaked is higher) than
our improved protocol for any value of the length $n$ of the strings being
compared.

(II) Comparing with our previous protocol in Ref. \cite{HeIJQI16}, as we
mentioned in section V.C, when taking $m\geq 14$, the WCWZ protocol is
always less secure than ours in Ref. \cite{HeIJQI16} for any length of the
strings being compared. Thus the WCWZ protocol may be valuable only when $%
2\leq m\leq 13$, which is covered by the area between the black solid line
and the black dashed line in Fig. 3. Even in this range, when $m=2$ ($m=13$%
), we can see that the amount of information leaked in the WCWZ protocol is
still higher than that in our previous one in Ref. \cite{HeIJQI16} for $n\leq 10$
($n\leq 60$). In fact, even our improved protocol in the current paper
(where $m=1$) cannot be less secure than our previous one when $n\leq 8$.
Furthermore, as we emphasized in section III, $I_{A}$ and $I_{B}$ in Fig. 3 are merely
the \textit{loose} bounds of the protocol in Ref. \cite{HeIJQI16}. Thus the claim in Ref. \cite%
{qi1696} that \textquotedblleft He's original protocol is not suitable for a
smaller bit-length comparison protocol\textquotedblright\ is obviously
wrong. Instead, the WCWZ protocol is even worst for comparing short strings.

(III) The simultaneity problem. Both step 7 of the WCWZ protocol and step iii
of our improved protocol require the existence of simultaneity. Thus our
previous protocol in Ref. \cite{HeIJQI16} wins again as it does not have this
requirement. In case simultaneously publishing informations is impossible,
both the WCWZ protocol and our improved one need modification. Then the
amount of information leaked in these two protocols could be even higher
than what is shown in Fig. 3.

%\bigskip

\subsection{Feasibility}

This is where our protocols really shine. As we stated in section V.D, the
WCWZ protocol not only requires the quantum memory for storing $2(k+2m)$
qubits, but also $2n$ pairs of Bell states for comparing two $n$-bit strings
$a$ and $b$.

On the contrary, in our previous protocol in Ref. \cite{HeIJQI16}, Alice and Bob
merely need to send qubits prepared in the pure states $\left\vert
0\right\rangle $, $\left\vert 1\right\rangle $, $\left\vert +\right\rangle $
and $\left\vert -\right\rangle $\ non-entangled with other systems. Also, no
quantum memory is required because once Alice and Bob receive the qubits,
they can measure them immediately without the need to wait for the other
party to announce further information.

Similarly, our improved protocol has the same low requirement on quantum resource
too. This is because only step i is quantum as it involves a QKD process.
All other steps are completely classical. As it is known, there exist QKD
protocols which can be run without entangled states and quantum memory. The
original BB84 protocol Ref. \cite{qi365} is exactly such an example.

Therefore, our two protocols are both much more feasible than the WCWZ
protocol, and can be implemented with currently available technology.
Especially, the improved protocol proposed in the current paper can be
realized directly using existing QKD systems.

\section{Summary}

We analyzed the amount of information leaked in the WCWZ protocol, and found
that it is less secure than our previous protocol in Ref. \cite{HeIJQI16} against
internal attacks when $m\geq 14$. For comparing short bit-strings with the
length $n\leq 10$, the WCWZ protocol is always less secure than ours no
matter which $m$ value it chooses, in contrast to the claim in Ref. \cite{qi1696}.

We also proposed an improved protocol, which is more secure than the WCWZ
protocol for any length of the strings being compared. Moreover, the WCWZ
protocol has to rely on the use of quantum memory and entanglement, while
these resources are not needed in both of our improved protocol and the
previous one.

\section*{Acknowledgements}

%\ack

The work was supported in part by Guangdong Basic and Applied Basic Research
Foundation under grant No. 2019A1515011048.

%China.

%\appendix

\section*{Appendix}

According to Ref. \cite{qi134}, if simultaneity is available, then perfectly
secure coin flipping (CF) a.k.a. coin tossing can be achieved.

CF is aimed to provide a method for two separated parties Alice and Bob to
generate a random bit value $c=0$ or $1$ remotely, while they do not trust
each other. A CF protocol is considered secure if neither party can bias the
outcome, so that $c=0$ and $c=1$ will both occur with the equal probability $%
1/2$, just as if they are tossing an ideal fair coin.

Assume that there is an approach to ensure both Alice and Bob to send
messages simultaneously. The CF protocol in Ref. \cite{qi134} can be
described in plain words as follows.

\bigskip

\textit{CF Protocol:}

I) Alice picks a random bit $a\in \{0,1\}$ and Bob picks a random bit $b\in
\{0,1\}$.

II) Alice and Bob publish $a$ and $b$ simultaneously.

III) They take $c=a\oplus b$ as the result of the protocol.

\bigskip

The simultaneous publishing of information in step II can be done using the approach described in the second paragraph of section V.B, where both Alice and Bob should have agents to test and confirm that they are located in the prescribed regions. This enables them to deduce the sending time of the other party from the time they receive the other's published information, so that none of them can delay the publishing of information without being discovered. This step is
the key to the security of the protocol. It guarantees that as
long as the two parties are separated spatially, by the time when Alice is forced to publish
her bit $a$, she has not received Bob's bit $b$ yet because the latter information has to spend a non-vanishing period of time to travel the distance between Alice and Bob due to the impossibility
of superluminal signaling. Consequently, the possible value of $b$ is unknown and looks
completely random to Alice at this stage, so that no matter she publishes $a=0$
or $a=1$, the final result $c$ can be $0$ or $1$ with equal probability $1/2$%
. Bob's case is exactly the same. Therefore, the bias in this protocol is
absolutely zero \cite{qi134}. Thus we can see that the existence of
simultaneity can easily evade the no-go proofs of nonrelativistic
unconditionally secure CF with an arbitrarily small bias \cite%
{qi58,qi76,qi817,qi145,qbc37,qi246,qbc132,qbc19}.

%\section*{References}

\end{document}